\documentclass[preprint,superscriptaddress,aps]{revtex4}
\usepackage{amsfonts}
\usepackage{graphics}
\usepackage{graphicx}
\usepackage{slashed}
\newcommand{\be}{\begin{equation}}
\newcommand{\ee}{\end{equation}}
\newcommand{\en}{\end{equation}}
\newcommand{\ba}{\begin{eqnarray}}
\newcommand{\ea}{\end{eqnarray}}
\newcommand{\bea}{\begin{eqnarray}}
\newcommand{\eea}{\end{eqnarray}}

\def\pls{\partial\!\!\!/}

\def\ps{p\!\!\!/}

\def\bs{b\!\!\!/}

\begin{document}

\title{On the higher-derivative Lorentz-breaking terms}
\thanks{Invited Contribution to the Proceedings of the 7th International Conference on Mathematical Methods in Physics,
Rio de Janeiro, Brazil, April 16–20, 2012; to be published in Proceedings of Science.}
\author{T. Mariz}
\affiliation{Instituto de F\'\i sica, Universidade Federal de Alagoas, 57072-270, Macei\'o, Alagoas, Brazil}
\email{tmariz@fis.ufal.br}
\author{J. R. Nascimento}
\affiliation{Departamento de F\'{\i}sica, Universidade Federal da Para\'{\i}ba\\
 Caixa Postal 5008, 58051-970, Jo\~ao Pessoa, Para\'{\i}ba, Brazil}
\email{jroberto, petrov@fisica.ufpb.br}
\author{A. Yu. Petrov}
\affiliation{Departamento de F\'{\i}sica, Universidade Federal da Para\'{\i}ba\\
 Caixa Postal 5008, 58051-970, Jo\~ao Pessoa, Para\'{\i}ba, Brazil}
\email{jroberto, petrov@fisica.ufpb.br}



\begin{abstract}
The first one-loop higher-derivative contribution to the effective action of the Lorentz-breaking spinor QED is obtained and shown to be finite and ambiguous.
\end{abstract}

\maketitle

It is well known that the Lorentz symmetry breaking essentially increases the possibilities for the extension of the field theory models by the new additive terms \cite{Kostel,Kostel2,Kostel3}. The most natural classification of such terms is based on the derivative expansion \cite{DE,DE2,DE3,DE4}. In the case of the Lorentz-breaking extension of the spinor QED which we will discuss here, the only restrictions on these modifications arises from the requirement of their gauge invariance. The first term studied within this context is the Carroll-Field-Jackiw (CFJ) term \cite{CFJ} involving only one derivative. The next term arising in the derivative expansion is the aether-like term  \cite{aether,aether2,aether3}. Thus, the natural continuation of study of the one-loop effective action of the Lorentz-breaking spinor QED consists in discussion of the higher-derivative contribution. The example of such a term is the Myers-Pospelov term, involving three derivatives, known for the fact that it generates the rotation of the polarization plane of light \cite{MP}. Beside of this term, the higher-derivative CFJ term different from the usual CFJ term by the presence of the extra d'Alembertian operator also can be generated. In this paper, we consider the perturbative generation of these terms.

Our starting point is the Lorentz-breaking extension of the spinor electrodynamics involving two couplings, that is, minimal one, proportional to $e$, and the noniminimal one, proportional to $g$ \cite{ourhd}: 
\begin{equation}
\label{mcn}
{\cal L}=\bar{\psi}\left[i \pls- \gamma^{\mu}(eA_{\mu}+g\epsilon_{\mu\nu\lambda\rho}F^{\nu\lambda}b^{\rho}) - m -  \gamma_{5}\bs\right]\psi-\frac{1}{4}F_{\mu\nu}F^{\mu\nu}.
\end{equation}
Here, the $b_\rho$ is a constant vector implementing the Lorentz symmetry breaking, and $F_{\mu\nu}=\partial_{\mu}A_{\nu}-\partial_{\nu}A_{\mu}$ is the usual stress tensor constructed on the base of the gauge field $A_{\mu}$. 

The corresponding one-loop effective action of the gauge field $A_{\mu}$, which we will denote as  $S_{eff}[b,A]$, can be expressed in the form of the following functional trace:
\begin{equation}
\label{det}
S_{eff}[b,A]=-i\,{\rm Tr}\,\ln(\ps-\gamma^{\mu}\tilde{A}_{\mu}- m - \gamma_5 \bs),
\end{equation}
where 
\begin{eqnarray}
\label{tilde}
\tilde{A}_{\mu}=eA_{\mu}+g\epsilon_{\mu\nu\lambda\rho}F^{\nu\lambda}b^{\rho}.
\end{eqnarray}
The $S_{eff}[b,A]$ can be expanded in the following power series,
\begin{equation}
\label{ea}
S_{eff}[b,A]=i\,{\rm Tr} \sum_{n=1}^{\infty}\frac1n
\Biggl[\frac1{\ps- m - \gamma_5 \bs}\, \gamma^{\mu}\tilde{A}_{\mu}\Biggr]^n.
\end{equation}
Within this study, we are going to consider the contributions of the second order in $\tilde{A}_{\mu}$ (further we will restrict ourselves only by the terms of the third order in $b_{\mu}$). The relevant expression is given by
\begin{equation}
S_{eff}^{(2)}[b,A]=\frac{i}{2}{\rm Tr}\frac{1}{ \ps- m - \gamma_5 \bs}\;\gamma^{\mu}\tilde{A}_{\mu}\;\frac{1}{ \ps- m - \gamma_5 \bs}\;\gamma^{\nu}\tilde{A}_{\nu},
\end{equation}
or, as is the same,
\begin{equation}
S_{eff}^{(2)}[b,A]=\frac{i}{2}\int d^4x\, \Pi_b^{\mu\nu}\tilde{A}_{\mu}\tilde{A}_{\nu},
\end{equation}
where
\begin{equation}
\label{Pib}
\Pi_b^{\mu\nu}={\rm tr}\int\frac{d^4p}{(2\pi)^4}\frac{1}{\ps-m-\gamma_5\bs }\gamma^\mu\frac{1}{\ps-i\pls -m-\gamma_5\bs }\gamma^\nu.
\end{equation}
From these equations, one can find that the contribution to the one-loop effective action of the third order in $b_{\mu}$ is given by three terms, where the number of insertions of the vector $b^{\mu}$ into the propagators is equal to one, two or three. This corresponds to two, one or zero vertices of the form $g\bar{\psi}\epsilon_{\mu\nu\lambda\rho}\gamma^{\mu}F^{\nu\lambda}b^{\rho}\psi$, respectively (further we will refer to these vertices as to nonminimal ones). Let us consider all these situations.

First, we consider a contribution to $\Pi_b^{\mu\nu}$ characterized by two nonminimal vertices. It follows from (\ref{Pib}) that it is given by 
\begin{eqnarray}
\label{I1}
\Pi_{b1}^{\mu\nu}(k) &=& {\rm tr}\int\frac{d^4p}{(2\pi)^4}S(p)\gamma_5\bs S(p)\gamma^\mu S(p-k)\gamma^\nu \nonumber\\
&&+{\rm tr}\int\frac{d^4p}{(2\pi)^4}S(p)\gamma^\mu S(p-k)\gamma_5\bs S(p-k)\gamma^\nu,
\end{eqnarray}
with $S(p)=(\ps-m)^{-1}$. The same trace emerges when the CFJ term is perturbatively generated (for more details, see f.e. \cite{ourLV,ourLV2} and references therein). The result, up to the first order in the external momentum $k_{\mu}$, is $\Pi_{b1}^{\mu\nu} =C\epsilon^{\mu\nu\lambda\rho}b_\lambda k_\rho$, where the coefficient $C$, being formally superficially divergent, has been shown to be finite and ambiguous, depending essentially on the regularization scheme (for the discussion on the ambiguity see f.e. \cite{JackAmb}). 

The corresponding contribution to the effective action can be written as
\begin{equation}
\label{term1}
S_{b1}=2g^2C\int d^4x\,  \left[b^\alpha F_{\alpha\mu}(b\cdot\partial)b_\beta\epsilon^{\beta\mu\nu\lambda}F_{\nu\lambda}+b^2b_\beta\epsilon^{\beta\mu\nu\lambda}A_\mu\Box F_{\nu\lambda}\right].
\end{equation}
The first term of this expression is the Myers-Pospelov term \cite{MP}, and the second one is the higher-derivative CFJ term (earlier discussed in \cite{Mariz,Mariz2}). Both these terms are gauge invariant. 

Now, let us consider another two contributions of third order in $b^{\mu}$, which essentially involve the minimal coupling. First, we evaluate the contribution to the effective action involving one vertex with minimal coupling, and another vertex with nonminimal coupling. The corresponding contribution from Eq.~(\ref{Pib}) is given by
\begin{eqnarray}
\Pi_{b2}^{\mu\nu} &=& {\rm tr}\int\frac{d^4p}{(2\pi)^4}S(p)\gamma_5\bs S(p)\gamma_5\bs S(p)\gamma^\mu S(p-k)\gamma^\nu \nonumber\\
&&+{\rm tr}\int\frac{d^4p}{(2\pi)^4}S(p)\gamma_5\bs S(p)\gamma^\mu S(p-k)\gamma_5\bs S(p-k)\gamma^\nu \nonumber\\
&&+{\rm tr}\int\frac{d^4p}{(2\pi)^4}S(p)\gamma^\mu S(p-k)\gamma_5\bs S(p-k)\gamma_5\bs S(p-k)\gamma^\nu.
\end{eqnarray}
After calculating the trace and taking into account only the terms with the third derivative, we arrive at
\begin{eqnarray}
\label{term2}
S_{b2}=\frac{eg}{6\pi^2m^2}\int d^4x\, \left[b^\alpha F_{\alpha\mu}(b\cdot\partial)b_\beta\epsilon^{\beta\mu\nu\lambda}F_{\nu\lambda}+b^2b_\beta\epsilon^{\beta\mu\nu\lambda}A_\mu\Box F_{\nu\lambda}\right],
\end{eqnarray}
We see that in this case, again both the Myers-Pospelov term and the higher-derivative CFJ term arise.

It remains to consider only the contribution to the effective Lagrangian involving both vertices with minimal coupling which requires three insertions of the $\gamma_5\bs$ into the propagator. Thus, we have
\begin{eqnarray}
\Pi_{b3}^{\mu\nu} &=& {\rm tr}\int\frac{d^4p}{(2\pi)^4}S(p)\gamma_5\bs S(p)\gamma_5\bs S(p)\gamma_5\bs S(p)\gamma^\mu S(p-k)\gamma^\nu \nonumber\\
&&+{\rm tr}\int\frac{d^4p}{(2\pi)^4}S(p)\gamma_5\bs S(p)\gamma_5\bs S(p)\gamma^\mu S(p-k)\gamma_5\bs S(p-k)\gamma^\nu \nonumber\\
&&+{\rm tr}\int\frac{d^4p}{(2\pi)^4}S(p)\gamma_5\bs S(p)\gamma^\mu S(p-k)\gamma_5\bs S(p-k)\gamma_5\bs S(p-k)\gamma^\nu \nonumber\\
&&+{\rm tr}\int\frac{d^4p}{(2\pi)^4}S(p)\gamma^\mu S(p-k)\gamma_5\bs S(p-k)\gamma_5\bs S(p-k)\gamma_5\bs S(p-k)\gamma^\nu.
\end{eqnarray}
After calculating the trace, integration over momenta and the Fourier transform, we obtain
\begin{eqnarray}
\label{term3}
S_{b3}=\frac{4e^2}{45\pi^2m^4}\int d^4x\, \left[b^\alpha F_{\alpha\mu}(b\cdot\partial)b_\beta\epsilon^{\beta\mu\nu\lambda}F_{\nu\lambda}+\frac54b^2b_\beta\epsilon^{\beta\mu\nu\lambda}A_\mu\Box F_{\nu\lambda}\right].
\end{eqnarray}
This term is superficially finite and free of any ambiguities, as well as (\ref{term2}). 

We see that the three-derivative contribution to the self-energy tensor arisen in the theory is given by the sum of the expressions (\ref{term1}), (\ref{term2}), and (\ref{term3}). Its explicit form is
\begin{eqnarray}
\label{term}
S_{hd}&=&\left(2g^2C+\frac{eg}{6\pi^2m^2}+\frac{4e^2}{45\pi^2m^4}\right)\int d^4x\, b^\alpha F_{\alpha\mu}(b\cdot\partial)b_\beta\epsilon^{\beta\mu\nu\lambda}F_{\nu\lambda}\nonumber\\
&&+\left(2g^2C+\frac{eg}{6\pi^2m^2}+\frac{e^2}{9\pi^2m^4}\right)\int d^4x\,b^2b_\beta\epsilon^{\beta\mu\nu\lambda}A_\mu\Box F_{\nu\lambda}.
\end{eqnarray}
We see that this contribution is finite and gauge invariant. It is interesting to point out that for the light-like Lorentz-breaking vector $b^{\mu}$, the higher-derivative CFJ term disappears. However, we note that the light-like $b^{\mu}$ leads to instabilities \cite{Reyes}, whereas the space-like $b^{\mu}$ is preferable. This is also consistent with the analysis related to the minimal Lorentz-breaking spinor QED \cite{Lehnert}, since the time-like $b^{\mu}$ can produce a certain stability problem, although it is not clear whether this problem affects the consistency of the theory.  Therefore, it is natural to suggest that the higher-derivative CFJ term must present in a consistent theory.

We have carried out the perturbative generation of the three-derivative gauge-invariant term in the Lorentz-breaking extended spinor QED. This term represents itself as a linear combination of the Myers-Pospelov term and the higher-derivative CFJ term, being finite, gauge invariant and ambiguous. The ambiguity of the coefficient $C$ accompanying this term is identically the same as that one accompanying the CFJ term in the usual Lorentz-breaking QED \cite{ourLV,ourLV2}. This shows that the ABJ anomaly which is known to be related with the ambiguity of the CFJ term \cite{JackAmb} can be naturally promoted to the higher-derivative theories. One must note, however, that this ambiguity disappears if we switch off the nonminimal interaction, therefore, there is no this ambiguity in the usual Lorentz-breaking QED, although the Myers-Pospelov term arises even in this case as we have shown. 

{\bf Acknowledgments.} This work was partially supported by Conselho
Nacional de Desenvolvimento Cient\'{\i}fico e Tecnol\'{o}gico (CNPq). The work by A. Yu. P. has been supported by the
CNPq project No. 303461/2009-8.

\end{document}